\documentclass[aps,prb,reprint,superscriptaddress,footinbib,floatfix]{revtex4-2}

\usepackage{amsmath,amssymb,bm}
\usepackage{graphicx}
\usepackage{xcolor}
\usepackage[colorlinks=true,linkcolor=blue,citecolor=blue,urlcolor=blue]{hyperref}

\usepackage{titlesec}

\begin{document}

\title{Transverse quasiparticle heat transport in
\texorpdfstring{$p$}{p}-wave magnetic Josephson junctions}

\author{Jin-Xing Hou }
%\thanks{These authors contributed equally.}
\affiliation{Hefei National Laboratory, Hefei, Anhui, 230088, China}
\affiliation{International Center for Quantum Design of Functional Materials (ICQD), University of Science and Technology of China, Hefei, Anhui 230026, China}

\author{Pritam Chatterjee}
%\thanks{These authors contributed equally.}
\affiliation{Hefei National Laboratory, Hefei, Anhui, 230088, China}
\affiliation{International Center for Quantum Design of Functional Materials (ICQD), University of Science and Technology of China, Hefei, Anhui 230026, China}

\author{Song-Bo Zhang}
\email{songbozhang@ustc.edu.cn}
\affiliation{Hefei National Laboratory, Hefei, Anhui, 230088, China}
\affiliation{International Center for Quantum Design of Functional Materials (ICQD), University of Science and Technology of China, Hefei, Anhui 230026, China}

% \author{et al.}

% Add the remaining authors and affiliations before submission.

\date{\today}

\begin{abstract}
We study transverse heat currents carried by Bogoliubov quasiparticles in $p$-wave magnetic Josephson junctions under a longitudinal temperature bias. We find that such a junction supports a transverse spin heat current while the net transverse heat current vanishes. This effect stems from the $p$-wave magnetism, which shifts the two spin sectors oppositely in transverse momentum space and thereby causes them to carry opposite heat currents.    
A weak Zeeman field lifts the compensation between the two spin-sector contributions, thereby generating a finite net transverse heat current. The directions of both currents can be switched by reversing the sign of the $p$-wave magnetic splitting or tuning the superconducting phase difference, while both currents vanish in the absence of $p$-wave magnetism. These findings highlight $p$-wave magnetic Josephson junctions as a versatile platform for switchable transverse heat transport and controllable spin caloritronics.

\end{abstract}

\maketitle

\textit{\color{blue}Introduction}.\textemdash 
Unconventional magnetism has recently emerged as a vibrant area of research in condensed-matter physics~\cite{Smejkal2022PRX011,Smejkal2022PRX031,Smejkal2022PRX040,Krempasky2024,Fedchenko2024}. Such systems can host sizable spin splitting despite negligible net magnetization, opening new avenues for spin-dependent transport and functional quantum devices~\cite{Ouassou2023,Beenakker2023,Papaj2023,Sun2023,Sun2023b,Zhang2024,Tanaka-PRL-2024,Tanaka-JPSJ-2024, Hodt2024, Hedayati2025,Jin-Xing2026, Jin-Xing2026II}. 
Of particular interest are $p$-wave magnets (PMs), which form a distinct class of odd-parity magnetic systems characterized by momentum-antisymmetric spin splitting~\cite{Brekke2024,Sukhachov2024,Chakraborty2025,Ezawa2025,Law2025,Ezawa2025PRB,ZhuangArxiv2025,Leeb2026, Sim2026,Khodas2026,Šmejkal2026,mitscherling2026,li2026pwaveorbitalmagnetism,LiuPRB2026}, in contrast to the momentum-even spin splitting of conventional ferromagnets and altermagnets. Recently, several materials have been predicted to host $p$-wave magnetism~\cite{hellenes2024pwavemagnets,León2025,Sivianes2025,McNally2026,zhang2026,zhou2026,eto2026,wang2026,Hongming2026,cuono2026,jeong2026}, and experimental evidence for this magnetic state has also been reported~\cite{Yamada2025,Song2025}. Notably, the odd-parity splitting preserves the energy matching between opposite-momentum, opposite-spin states~\cite{Brekke2024,Hodt2025}. This property makes PMs compatible with conventional spin-singlet superconductivity, motivating the study of PM-based superconducting (SC) junctions~\cite{Maeda2025,Soori2025,Nagae2025,Sukhachov2025,Zeng2025,salehi2025,Tanaka2025,Wang2025,Bobkov2025,Yi2026,Cayao2026,Yang2026,Froldi2026}. 
% ~\cite{Hellenes2023,Brekke2024,Song2025,Yamada2025,Chakraborty2025}

The interplay between $p$-wave magnetism and superconductivity extends beyond Cooper-pair transport to thermally driven transport by Bogoliubov quasiparticles. Under a temperature bias, these above-gap excitations can propagate between the SC reservoirs and mediate dissipative heat flow~\cite{beenakker1991universal,brouwer1997anomalous,Zhao2003,Zhao2004,Giazotto2012}, in contrast to the nondissipative phase-coherent supercurrent carried by Cooper pairs. Previous studies have focused primarily on longitudinal heat transport across the junction~\cite{Yokoyama2008PRB,ghanbari2011wiedemann,Paul2016PRB,Sothmann_2016,Sothmann_2017,Dutta2017PRB,Savander2020PRR,Mukhopadhyay2021PRB,Acciai2021PRB,Mukhopadhyay2022,Chatterjee2023,Chatterjee_2024,Bera2025,Debnath2026}. 
By contrast, transverse quasiparticle heat transport parallel to the junction interfaces remains poorly understood, particularly regarding how odd-parity magnetic order and the SC phase difference govern the spin and net heat currents.

\begin{figure}[t]
\centering
\includegraphics[width=1\linewidth]{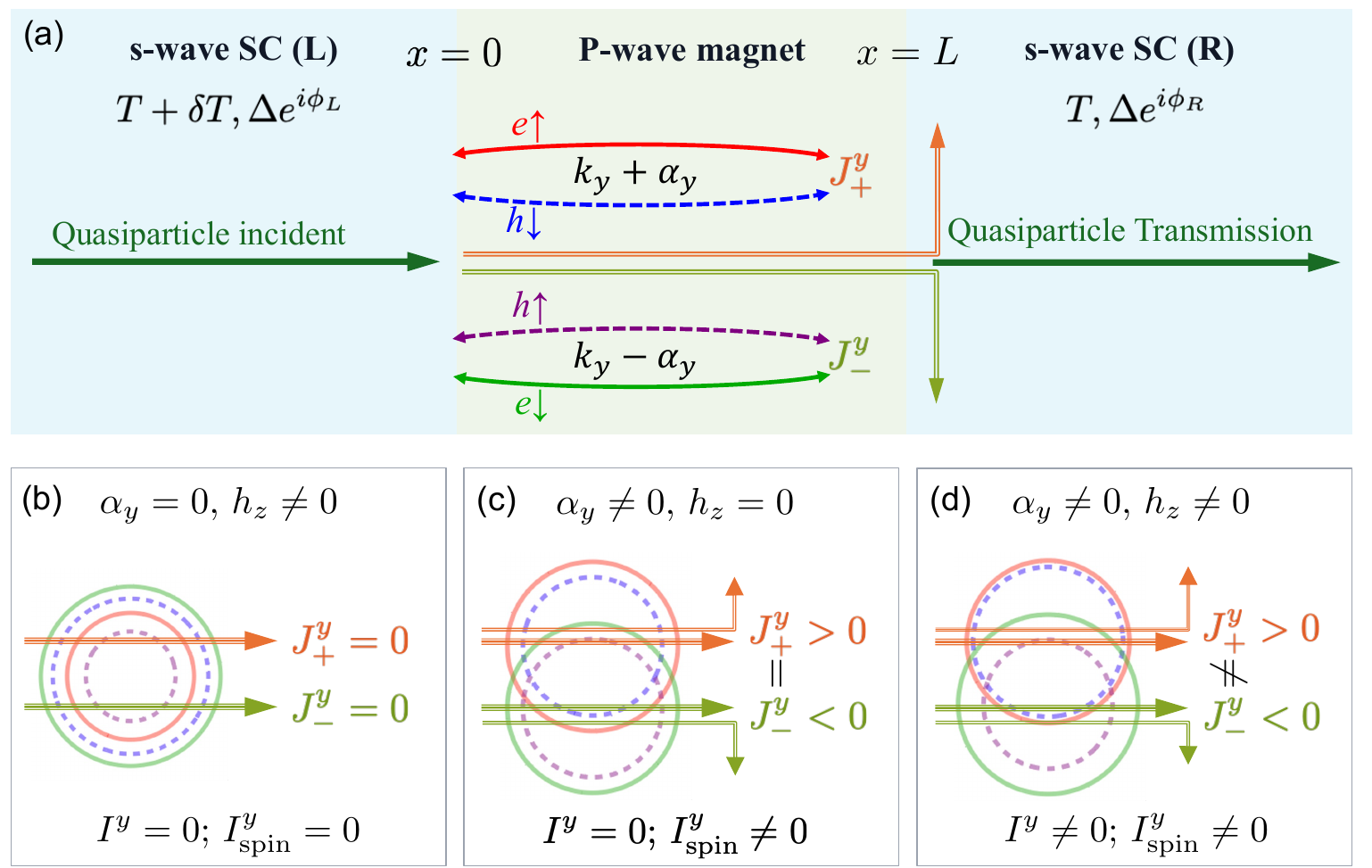 }
\caption{(a) Schematic of the SC/PM/SC junction with SC electrodes of phases $\phi_{L,R}$ under a temperature bias $\delta T$. The PM region of length $L$ exhibits spin-dependent transverse momentum shifts $\pm\alpha_y$ for the two spin sectors. 
(b)–(d) Spin- and particle-hole-resolved constant-energy contours in the PM region for (b) a Zeeman-split configuration, 
yielding no transverse heat current; (c) a pure PM configuration, 
supporting a transverse spin heat current; and (d) a PM state with Zeeman splitting,
yielding a finite transverse heat current. 
}
\label{fig1:Schematic}
\end{figure}

In this work, we investigate transverse quasiparticle heat transport in a planar Josephson junction consisting of a PM sandwiched between two conventional $s$-wave SC electrodes, using the Blonder-Tinkham-Klapwijk (BTK) formalism~\cite{Blonder1982}. We show that the PM splitting displaces the spin-resolved Fermi surfaces in opposite directions in transverse momentum space, producing distinct quasiparticle propagation channels for the two opposite-spin sectors. These two sectors carry opposite transverse heat currents, yielding a pure transverse spin heat current without accompanying net transverse heat flow, as illustrated in Figs.~\ref{fig1:Schematic}(b,c). A weak Zeeman field generates an imbalance between the two sectors, thereby generating a finite net transverse heat current [Fig.~\ref{fig1:Schematic}(d)]. 
We also identify the symmetry relations governing the currents and show that their directions can be reversed either by changing the sign of the PM splitting or by tuning the SC phase difference. These results reveal how $p$-wave magnetism, Zeeman field, and SC phase difference govern transverse quasiparticle heat transport.

\textit{\color{blue}Setup and formalism.}\textemdash
We consider an SC/PM/SC planar Josephson junction where a PM ($0<x<L$) connects two conventional $s$-wave SC electrodes occupying $x<0$ and $x>L$ [Fig.~\ref{fig1:Schematic}(a)]. We assume translational invariance along $y$-direction so that the transverse momentum $k_y$ remains a good quantum number.
The SC electrode $j\in \{L,R\}$ is described by the Bogoliubov-de Gennes (BdG) Hamiltonian
\begin{equation}
H_{\mathrm{SC},j}(\bm{k})=
\begin{pmatrix}
\xi_S(\bm{k})\sigma_0
&
\Delta_j(T_j) e^{i\phi_j} i\sigma_y
\\
-\Delta_j^\ast(T_j) e^{-i\phi_j} i\sigma_y
&
-\xi_S(\bm{k})\sigma_0
\end{pmatrix},
\label{eq:H_SC}
\end{equation}
in the Nambu basis $(c_{\bm{k}\uparrow}, c_{\bm{k}\downarrow}, c_{-\bm{k}\uparrow}^{\dagger}, 
c_{-\bm{k}\downarrow}^{\dagger})^T$. Here, $\xi_S(\bm{k})=M \bm{k}^2-\mu_S$ is the normal kinetic term with  $\mu_S$ the chemical potential. $\Delta_j(T_j)$ denotes the magnitude of the pairing
potential at temperature $T_j$ and $\phi_j$ is the SC phase. For $T<T_c$, we approximate the temperature dependence of the pairing amplitude by $\Delta_j(T_j)=\Delta\tanh(1.74\sqrt{{T_c}/{T_j}-1})$, where $\Delta$ is the zero-temperature pairing amplitude and $T_c$ is the SC critical temperature~\cite{Bardeen1957,Muhlschlegel1959,sm2026_thermal_transport}.
The matrices $\sigma_0$ and $\sigma_i$ ($i=x,y,z$) are the identity and Pauli matrices in spin space, respectively.
We define the SC phase difference as
$\phi \equiv \phi_L-\phi_R$, and set $T_L=T+\delta T$ and $T_R=T$, where $\delta T$ is a small applied temperature bias.

We model the PM in the long-wavelength limit by the effective Hamiltonian 
$h_{\text{p}}(\bm{k})=M(\bm{k}+ \sigma_z \bm{\alpha})^2$, where $\bm{\alpha}=(\alpha_x,\alpha_y)$~\cite{Tanaka-JPSJ-2024,Brekke2024,Hedayati2025}. The two spin-polarized bands are shifted oppositely by $\bm{\alpha}$ in momentum space. In Nambu space, the corresponding Hamiltonian is
\begin{equation}
H_{\mathrm{PM}}(\bm{k})=
\begin{pmatrix}
h_{\text{p}}(\bm{k}) - \mu_N \sigma_0 & 0\\
0 & -h_{\text{p}}(-\bm{k}) + \mu_N \sigma_0
\end{pmatrix},
\end{equation}
where $\mu_N$ is the chemical
potential of the PM. The two spin components are decoupled in the PM. When the PM is coupled to spin-singlet SCs, the transport problem decouples into two sectors with bases
$(c_{\bm{k}\uparrow},c^\dagger_{-\bm{k}\downarrow})^T$ and
$(c_{\bm{k}\downarrow},c^\dagger_{-\bm{k}\uparrow})^T$, labeled by $s=\pm$,
respectively. In the PM region, the electron and hole modes in each sector have Nambu spinors  $\eta_e=(1,0)^T$ and $\eta_h=(0,1)^T$, respectively, with corresponding longitudinal momenta
\begin{align}
k_{\nu s}^{\tau}&=-s\alpha_x+\tau Q_{\nu s},
\label{eq:prl_PMmomenta}
\end{align}
where $Q_{\nu s} =
\sqrt{({\mu_N+\zeta_\nu E})/{M}-(k_y+s\alpha_y)^2}$, $\nu\in\{e,h\}$, $\zeta_e=1$, and $\zeta_h=-1$. The index
$\tau=\pm1$ labels the two longitudinal branches, and 
$\tau=+1$ ($\tau=-1$) denotes right-moving (left-moving) electron modes, with the assignment reversed for hole modes.

As shown by Eq.~\eqref{eq:prl_PMmomenta}, the transverse shift $\alpha_y$ modifies the longitudinal momenta of the PM modes and shifts their propagating-mode windows in opposite directions for the two spin sectors. Since $k_y$ is conserved across the interfaces, these changes alter the matching of PM and SC modes at the same $k_y$, thereby modifying quasiparticle transmission through the junction, as discussed below. By contrast, the longitudinal component $\alpha_x$ shifts
$k_{\nu s}^{\tau}$ uniformly within each spin sector, introducing a common phase factor
$e^{-is\alpha_x x}$ for all modes in that sector. This factor contributes only an overall phase to the transmission amplitudes but leaves the transmission probabilities unchanged in the present spin-conserving junction~\cite{sm2026_thermal_transport}. We therefore set $\alpha_x=0$ in the discussion of the transmission probabilities and heat currents below.

{\color{blue}
\begin{figure}[t]
\centering
\includegraphics[width=0.98\linewidth]{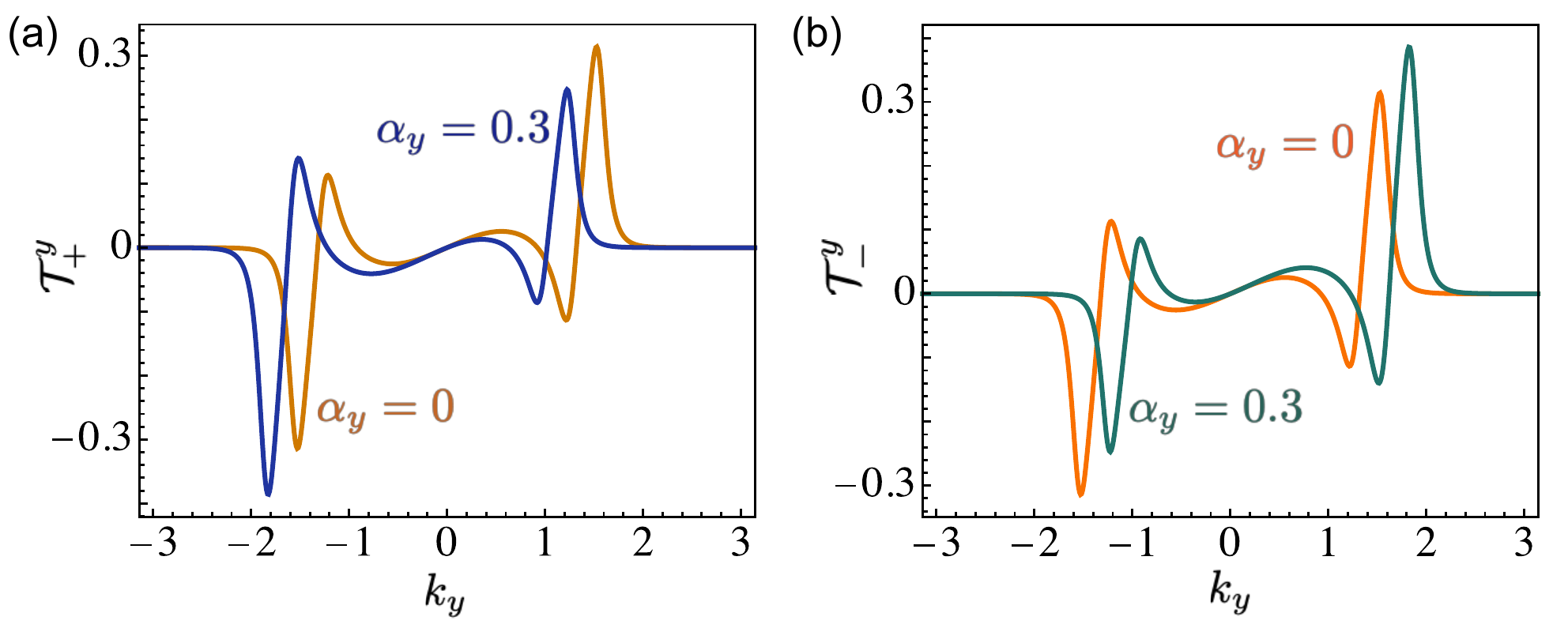}
\caption{Transverse heat-current kernel  
(a) $\mathcal{T}_+^y$ and (b) $\mathcal{T}_-^y$ as functions of transverse momentum $k_y$ at $E=2\Delta$ for $\alpha_y=0$ and $0.3$. Other parameters: $M=1$, $\Delta=0.2$, $k_B T_c=0.57\Delta$, $T=0.5T_c$, $\delta T=0.05T_c$, $L=3\xi_0$, $\mu_N=3\Delta$,
$\mu_S=20\Delta$, and $\phi=0.5\pi$, where
$\xi_0\equiv\sqrt{4M\mu_S}/\pi\Delta$ defines the SC coherence length. }
\label{fig:TransverseHeatKernel_ky}
\end{figure}
}

We focus on the transverse heat currents in the right SC driven by the temperature bias $\delta T$ applied across the junction. 
%{\color{red}~\footnote{For completeness, we present and discuss the longitudinal heat currents in the SM.}}. 
The heat current contributed by sector $s$ is given by~\cite{Scharf2016PRL,Linder2016,Bergeret2013,Ngampruetikorn2020,Uchida2021,sm2026_thermal_transport}
\begin{align}
J_s^y(\phi)
&=\frac{1}{h}
\int_{E_{\Delta}}^\infty\!\mathrm dE
\int_{-\infty}^\infty\,\frac{\mathrm dk_y}{2\pi}
E\,\mathcal T_s^y(E,k_y,\phi)\,\delta f(E),
\label{eq:sector_heatcurrent}
\end{align}
where $\mathcal T_s^y(E,k_y,\phi)$ is the transverse heat-current kernel whose explicit form is given below. Here, $\delta f(E)=f(E,T+\delta T)-f(E,T)$, with 
$f(E,T)=[\mathrm{exp}(E/k_B T)+1]^{-1}$ denoting the Fermi distribution function, and$E_{\Delta}=\max(\Delta_L,\Delta_R)$. 
The net transverse heat current is obtained by summing
the two sector contributions, whereas the transverse spin heat-current
density is defined by weighting them with their opposite spin
labels $s=\pm1$,
\begin{subequations}
\begin{align}
I_{\mathrm{net}}^y(\phi)
&=J_+^y(\phi)+J_-^y(\phi),
\label{eq:prl_heatcurrents}
\\
I_{\mathrm{spin}}^y(\phi)
&=J_+^y(\phi)-J_-^y(\phi).
\label{eq:prl_spinheatcurrents}
\end{align}
\end{subequations}
Since $\delta f(E)$ is independent of spin and transverse momentum, the spin and momentum dependence of $\mathcal T_s^y$ determines how the two sectors contribute to the transverse heat currents.

\textit{\color{blue}Transverse heat current kernels.}\textemdash
To determine the kernel $\mathcal T_s^y$, we calculate the scattering amplitudes within the BTK formalism by imposing continuity of the wavefunction $\Psi$ and $\hat v_x\Psi$ across the interfaces at $x=0$ and $x=L$. Here, $\hat v_x$ denotes the $x$ component of the velocity operator. These matching conditions ensure conservation of the quasiparticle flux normal to the interfaces. 
The transmission probabilities are obtained by normalizing the transmitted quasiparticle flux to the incident flux~\cite{Blonder1982,Zhao2003,Zhao2004}. Details of the derivation are provided in the SM~\cite{sm2026_thermal_transport}. 
Accounting for both electronlike and holelike incident channels, the transverse heat-current kernel in sector $s$ can be found as
\begin{align}
\mathcal T_{s}^{y}
=
\frac{k_y}{k_{e,R}}
\left(T_{ee}^{s}+T_{eh}^{s}\right)
-
\frac{k_y}{k_{h,R}}
\left(T_{he}^{\bar s}+T_{hh}^{\bar s}\right).
\end{align}
where $\bar s=-s$, $T_{e\nu}^{s}$ and $T_{h\nu}^{\bar s}$ denote the transmission probabilities into electronlike and holelike outgoing channels in the right SC, respectively, for a quasiparticle of type $\nu\in \{e,h\}$ incident from the left SC electrode. 
The longitudinal momenta in the right SC are given by $k_{\nu,R}= [(\mu_S+\zeta_\nu\Omega_R)/M-k_y^2]^{1/2}$, with $\Omega_R=\sqrt{E^2-\Delta_R^2}$, $\zeta_e=+1$ and $\zeta_h=-1$. 
The factors $k_y/k_{e,R}$ and $k_y/k_{h,R}$ correspond to the velocity ratios of the outgoing electronlike and holelike modes, respectively, converting longitudinal flux contributions into transverse ones. The relative minus sign arises from the opposite direction of the hole group velocity with respect to its wavevector~\cite{sm2026_thermal_transport}.

In the absence of a transverse PM shift ($\alpha_y=0$), the constant-energy contours in all regions are symmetric about $k_y=0$. Consequently, modes with opposite transverse momenta $\pm k_y$ encounter identical scattering conditions, making the transmission probabilities even in $k_y$. Since the transverse velocity weights are odd in $k_y$, the resulting kernel satisfies
$\mathcal T_s^y(E,k_y,\phi)=-\mathcal T_s^y(E,-k_y,\phi)$. A finite $\alpha_y$, however, shifts both the electronlike and holelike contours in the PM by $-s\alpha_y$ along $k_y$ for sector $s$. This displacement makes the scattering conditions at opposite transverse momenta inequivalent within the sector, so the kernels are generally no longer odd in $k_y$, as shown numerically in Fig.~\ref{fig:TransverseHeatKernel_ky}. %In the absence of external fields, 
The opposite displacements of the two sectors preserve the scattering symmetry under $(s,k_y)\to(-s,-k_y)$, yielding $\mathcal T_+^y(E,k_y,\phi)=-\mathcal T_-^y(E,-k_y,\phi)$. These modifications allow finite transverse heat currents in the individual sectors while constraining their relative signs, as discussed below.

\begin{figure}[t]
\centering
\includegraphics[width=0.98\linewidth]{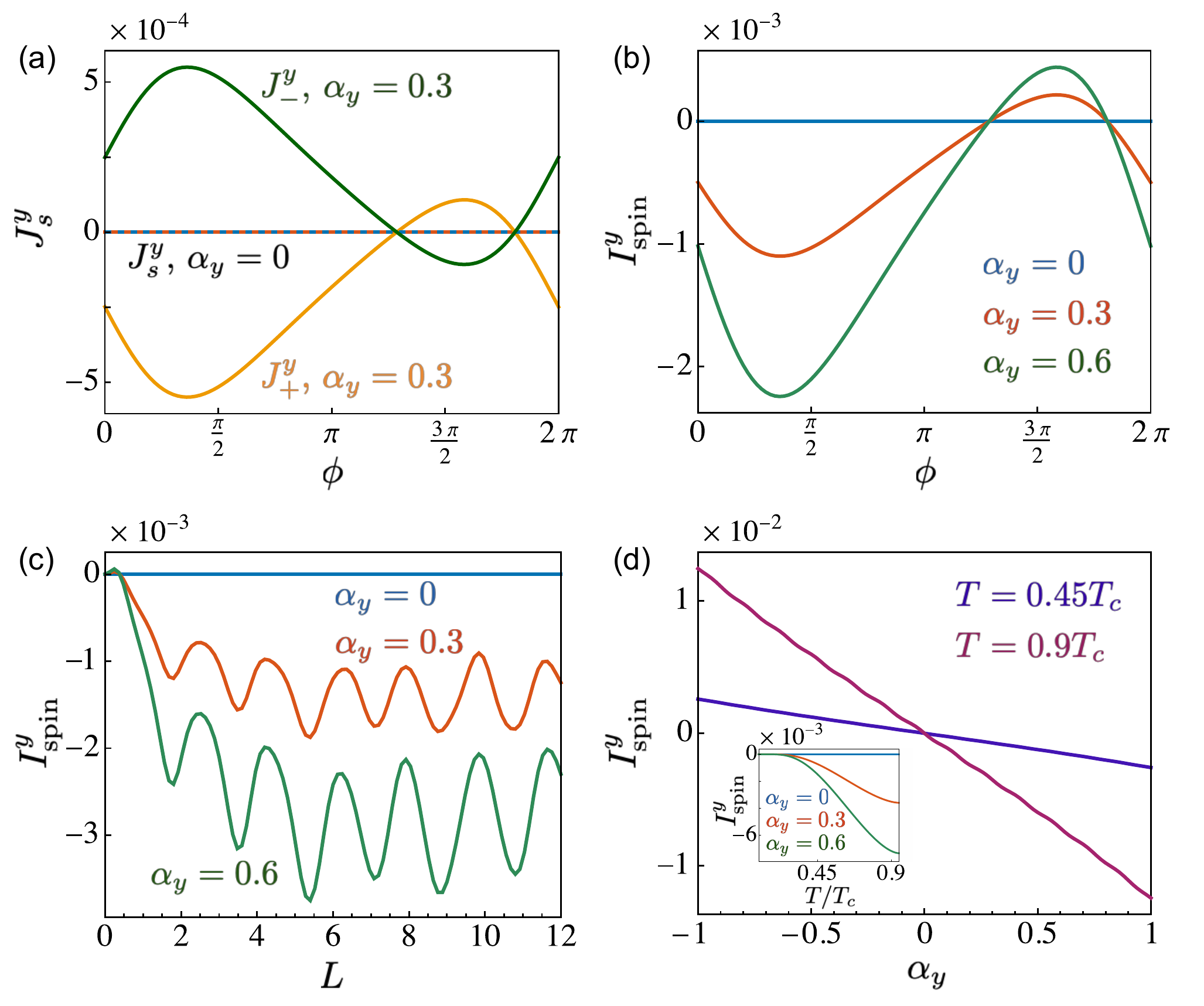}
\caption{(a) Sector-resolved transverse heat currents $J_{s}^{y}$ as functions of SC  phase difference $\phi$ for $\alpha_y=0$ (red) and $0.3$ (yellow and green). (b,c) Transverse spin heat current $I_{\mathrm{spin}}^{y}$ as a function of (b) $\phi$ and (c) the PM-region length $L$ (in units of $\xi_0$) at $\phi=0.5\pi$ for $\alpha_y=0$ (blue), $0.3$ (red), and $0.6$ (green). (d) $I_{\mathrm{spin}}^{y}$ as a function of PM splitting field $\alpha_y$ at $T=0.45T_c$ and $0.9T_c$, respectively. Inset: temperature dependence of $I_{\mathrm{spin}}^{y}$ for $\alpha_y=0$, 0.3 and 0.6, respectively. The currents are in units of $\Delta^2/(2\pi\hbar\xi_0)$. 
Other parameters are same as in Fig.~\ref{fig:TransverseHeatKernel_ky}. 
}
\label{fig:TransverseSpinHeatCurrent}
\end{figure}

\textit{\color{blue}Transverse spin heat currents}.\textemdash
For \(\alpha_y=0\), the transverse heat current vanishes in each sector, consistent with the odd parity of \(\mathcal T_s^y\) in \(k_y\).
A finite \(\alpha_y\) allows nonzero sector currents, while the intersector symmetry enforces $J_+^y(\alpha_y,\phi)=-J_-^y(\alpha_y,\phi)$, as illustrated in Fig.~\ref{fig:TransverseSpinHeatCurrent}(a). These equal and opposite contributions yield a pure transverse spin heat current,
\begin{align}
I_{\mathrm{net}}^y=0,\qquad
I_{\mathrm{spin}}^y=2J_+^y.
\end{align}
Moreover, for each sector, $J_s^y(-\alpha_y,\phi)=-J_s^y(\alpha_y,\phi)$. Hence,  $I_{\mathrm{spin}}^y$ reverses sign under $\alpha_y \rightarrow -\alpha_y$ and vanishes at $\alpha_y=0$. 
% Notably, the SC phase difference $\phi$ provides control over both the
% magnitude and direction of the spin heat current [Fig.~\ref{fig:TransverseSpinHeatCurrent}(b)]. {\color{red}(explain the zeros if know why they are independent of $\bm \alpha$). $\phi_0$-junction?} The current changes sign within a phase cycle and remains finite at $\phi=0$. These features reflect the distinct roles of the temperature bias and SC phase: the former drives heat transport, while the latter controls quasiparticle transmission. Increasing $\alpha_y$ enhances the phase dependence.
Notably, the SC phase difference $\phi$ controls both the magnitude and direction of the transverse spin heat current [Fig.~\ref{fig:TransverseSpinHeatCurrent}(b)]. 
At low temperatures, the current reverses sign near $\phi=3\pi/2$ at fixed temperature bias, while remaining finite at $\phi=0$. Increasing $\alpha_y$ enhances the phase modulation, whereas the zero-current phase positions vary only weakly over a broad range of $\alpha_y$, as shown in the SM~\cite{sm2026_thermal_transport}. These features reflect the distinct roles of the temperature bias and SC phase: the former drives heat transport, while the latter controls quasiparticle transmission.

To examine the role of propagation through the magnetic region, we calculate in 
Fig.~\ref{fig:TransverseSpinHeatCurrent}(c) $I_{\mathrm{spin}}^y$ as a function of the PM length $L$. For finite $\alpha_y$, the current exhibits pronounced oscillations, consistent with quasiparticle interference arising from propagation-phase accumulation in the magnetic region. Over most of the displayed range, varying $L$ modulates the
current magnitude without reversing its direction. A larger momentum shift produces stronger currents with more pronounced oscillations. 
%  while increasing $|\alpha_y|$ enhances its magnitude over the range considered. Moreover, {\color{red}the current at $T=0.9T_c$ is substantially stronger than at $T=0.45T_c$, indicating its temperature dependence.} {\color{red}The corresponding longitudinal heat currents are discussed in the SM~\cite{sm2026_thermal_transport}.}
Fig.~\ref{fig:TransverseSpinHeatCurrent}(d) illustrates the dependence on the transverse PM shift. In agreement with the symmetry analysis, reversing $\alpha_y$ switches the current direction, while increasing $|\alpha_y|$ enhances its magnitude over the range considered. Moreover, at fixed temperature bias, the current magnitude depends sensitively on the base temperature. As shown in the inset, $|I_{\mathrm{spin}}^y|$ increases monotonically with temperature in the low-temperature regime. 
However, the phase modulation weakens as $T$ approaches $T_c$, and phase-induced current reversal disappears. 
More details and the corresponding longitudinal heat currents are provided in the SM~\cite{sm2026_thermal_transport}.

\begin{figure}[t]
\centering
\includegraphics[width=1\linewidth]{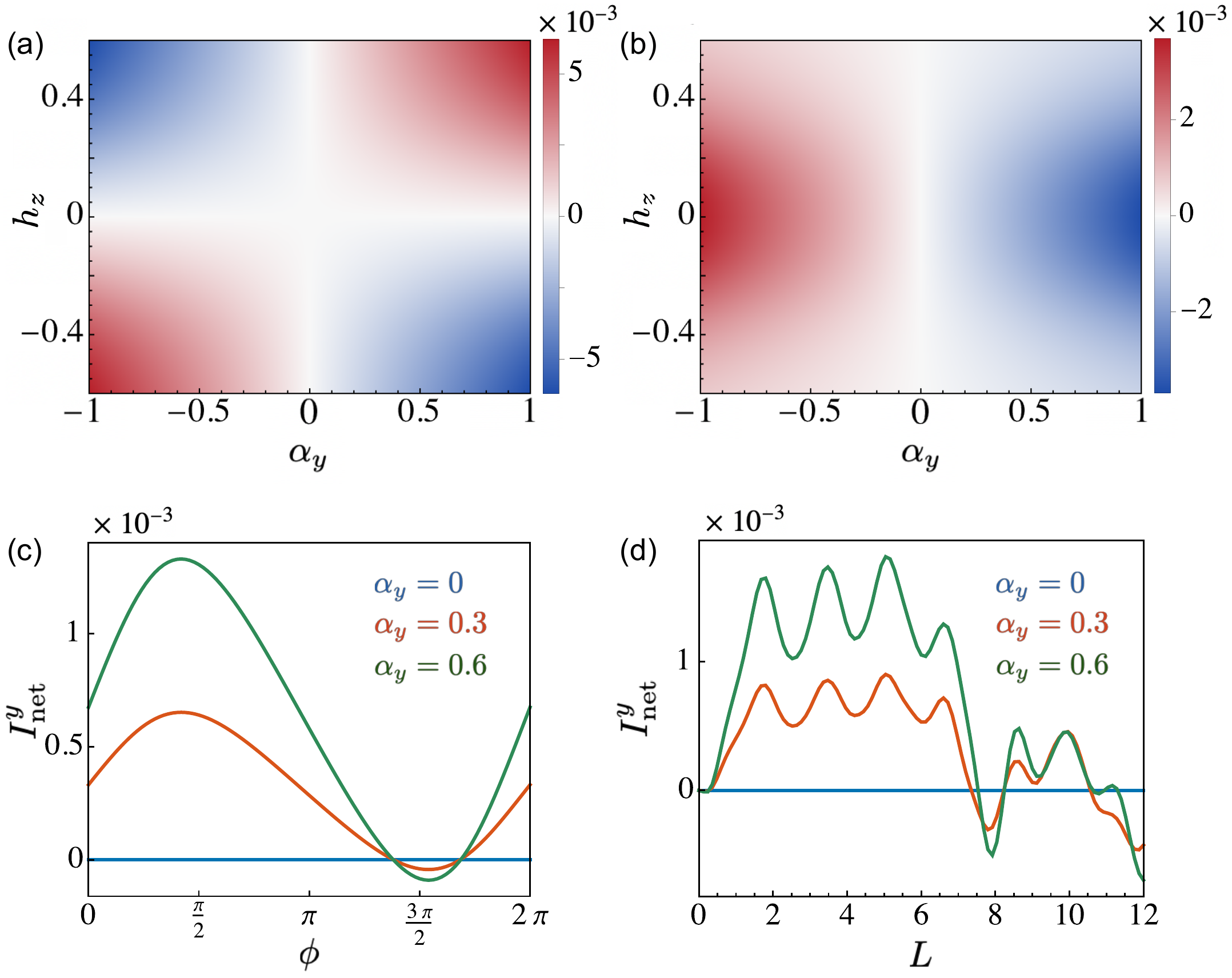}
\caption{Zeeman-field-controlled transverse heat currents. (a,b) Color maps of $I_{\mathrm{net}}^{y}$ (a) and $I_{\mathrm{spin}}^{y}$ (b) as functions of $\alpha_y$ and $h_z$. (c) Net transverse heat-current $I_{\mathrm{net}}^y$ as a function of $\phi$ at $h_z=0.2$ for $\alpha_y =0,0.3$ and $0.6$ respectively. Here, we consider junction length $L=3\xi_0$ for illustration.  (d) $I_{\mathrm{net}}^y$ as a function of $L$ at $h_z=0.2$ for $\alpha_y =0,0.3$ $0.6$ respectively. $\phi=0.5\pi$ in panels (a,b,d). 
Heat currents are in units of $\Delta^2/(2\pi\hbar\xi_0)$.
Other parameters:
$\Delta=0.2$, $k_B T_c=0.57\Delta$, $T=0.5T_c$,
$\delta T=0.05T_c$, $\mu_N=3\Delta$ and
$\mu_S=20\Delta$. 
}
\label{fig:transverse_heat}
\end{figure}

\textit{\color{blue}Zeeman-field controlled transverse heat transport}.\textemdash 
A net transverse heat current emerges when the compensation between
the two spin sectors is lifted. To achieve this, we introduce a
Zeeman term $H_B=h_z \sigma_z\tau_z$ in the PM region. This term leaves the two spin sectors decoupled but lifts the compensation between their transverse heat currents. Together with the transverse momentum shift $\alpha_y$, this enables a finite $I_{\mathrm{net}}^y$ when $\alpha_y\neq0$. 
Notably, the sector currents obey the antisymmetric
relations~\cite{sm2026_thermal_transport}:
\begin{subequations}
\begin{align}
J_{s}^{y}(-\alpha_y,h_z,\phi)=-J_{s}^{y}(\alpha_y,h_z,\phi),\\ J_{s}^{y}(\alpha_y,h_z,\phi)=-J_{-s}^{y}(\alpha_y,-h_z,\phi)
\end{align}
\end{subequations}
These lead to 
\begin{subequations}
\begin{align}
I_{\mathrm{net}}^y(-\alpha_y,h_z,\phi)
&=-I_{\mathrm{net}}^y(\alpha_y,h_z,\phi), \\
I_{\mathrm{net}}^y(\alpha_y,-h_z,\phi)
&=-I_{\mathrm{net}}^y(\alpha_y,h_z,\phi).
\label{eq:prl_transverse_heat_parity}
\end{align}
\end{subequations}
Thus, the net transverse heat current $I_{\mathrm{net}}^y$ requires both $\alpha_y$ and $h_z$ to be nonzero. Its direction reverses upon reversing either $\alpha_y$ or $h_z$ and remains unchanged when both are reversed. These properties are illustrated by the color map in Fig.~\ref{fig:transverse_heat}(a). The transverse spin heat current $I_{\mathrm{spin}}^y$, by contrast, remains finite at $h_z=0$
for nonzero $\alpha_y$ at the parameters shown
[Fig.~\ref{fig:transverse_heat}(b)]. It is odd in $\alpha_y$ but even in $h_z$. Thus, reversing $h_z$ switches $I_{\mathrm{net}}^y$ without changing $I_{\mathrm{spin}}^y$, whereas reversing the transverse PM shift switches both.

In addition to magnetic control, the net heat current can be tuned through the SC phase difference $\phi$ and junction length $L$. Figure~\ref{fig:transverse_heat}(c) shows $I_{\mathrm{net}}^y$ as a function of $\phi$ at $h_z=0.2$. For finite $\alpha_y$, the current remains finite at $\phi=0$ and reverses sign near $\phi=3\pi/2$ at low temperatures, demonstrating phase control of the thermally driven response.
Increasing $\alpha_y$ enhances the phase modulation, similar to the behavior of $I_{\mathrm{spin}}^y$. The sign reversal persists over a broad range of PM strengths, 
although the phase differences at which the net current changes sign oscillate with $\alpha_y$, whereas those of the spin heat current remain nearly unchanged (see SM~\cite{sm2026_thermal_transport} for more details).
Figure~\ref{fig:transverse_heat}(d) further shows pronounced oscillations with the PM length $L$, due to quasiparticle
interference in the magnetic region. Unlike the spin heat current in
Fig.~\ref{fig:TransverseSpinHeatCurrent}(c), the net heat current
also changes sign as $L$ varies over the displayed range. Thus, magnetic reversal, the SC phase difference, and the junction length provide distinct means of controlling the
direction of $I_{\mathrm{net}}^y$.

\textit{\color{blue}Summary and discussion}.\textemdash
To summarize, we have demonstrated that PM Josephson junctions support thermally driven transverse quasiparticle heat transport. The PM-induced momentum displacement generates
heat currents of equal magnitude in opposite directions in the two spin-polarized sectors, yielding a pure transverse spin heat current without net transverse heat flow. A Zeeman field lifts this compensation and enables finite net transverse heat currents. 
Both currents can be tuned through the SC phase difference, junction length, and Zeeman field. At low temperatures, varying the SC phase difference can reverse both currents. 
These results highlight PM Josephson junctions as a platform for switchable transverse heat transport and controllable spin caloritronics.

Recently, there have been an increasing number of materials proposed to host $p$-wave magnetism~{\cite{hellenes2024pwavemagnets,Chakraborty2025,Šmejkal2026}. Experimental evidence has also been reported in multiferroic $\mathrm{NiI_2}$~\cite{Song2025}, metallic systems including $\mathrm{CeNiAsO}$~\cite{zhou2026} and $\mathrm{Gd_3(Ru_{0.95}Rh_{0.05})_4Al_{12}}$~\cite{Yamada2025}. 
Electrical switching in $\mathrm{NiI_2}$ and field-driven domain control in $\mathrm{CeNiAsO}$ further demonstrate routes to manipulating the underlying magnetic order~{\cite{Song2025,zhou2026}.  
Odd-parity spin splitting preserves the energy matching between opposite-momentum, opposite-spin states, indicating symmetry compatibility with conventional spin-singlet pairing~\cite{Brekke2024}. 
Integrating these metallic PMs with SC electrodes such as Nb, Al, or NbSe$_2$ may therefore provide a route to testing our predictions. 

The transverse heat currents in response to the applied longitudinal temperature bias could be detected using lateral thermometry~\cite{Kim2019} or thermally monitored side contacts~\cite{Timossi2018}. Heat transport in Josephson interferometry~\cite{Giazotto2012} and transverse thermal measurements in kagome superconductor CsV$_3$Sb$_5$~\cite{Yoshida2025} have been experimentally investigated. 
The spin heat response requires spin-sensitive detection, potentially using spin-selective tunnel probes, motivated by recent experiments on spin heat accumulation and spin-dependent quasiparticle energy transport~\cite{Dejene2013,Kuzmanovic2020,Heidrich2019}.

\textit{\color{blue}Acknowledgments}.\textemdash
%We thank XXX for valuable discussions. 
This work was supported by the National Natural Science Foundation of China (Grant No. 12488101), HFNL Self-Deployed Project (Grant No. ZB2602000300), and Quantum Science and Technology-National Science and Technology Major Project (Grant No. 2021ZD0302801). 
% \end{acknowledgments}

\bibliography{arxiv_261105}

\end{document}